\documentclass[researcharticle,9pt,twoside,oneauthor]{Definitions/preprint}
\usepackage{epstopdf}
\usepackage{tabularx,array}
\newcolumntype{L}{>{\raggedright\arraybackslash}X}
\newcolumntype{Z}{>{\centering\arraybackslash}X}
\runninghead{Leakage-aware Hybrid acoustic and pretrained embedding feature fusion for UAV detection}

\Title{Leakage-aware acoustic drone detection: fusing hand-crafted and pretrained audio features in a stacking ensemble}
\Author{Zahid Iqbal\textsuperscript{1,}*}
\AuthorNames{Zahid Iqbal}
\AuthorCitation{Iqbal, Z.}
\address{%
	\textsuperscript{1}\quad Department of Computer Science, Faculty of Computing and Information Technology, University of Gujrat, 50700, Punjab, Pakistan}
\corres{* Corresponding author: Zahid Iqbal}

\abstract{\textbf{Introduction:} Small multirotor drones produce a characteristic acoustic signature, and microphones offer a low-cost way to detect them for airspace safety and drone-traffic monitoring. High accuracies have been reported on public drone audio benchmarks, often with random clip-level splits of datasets built from a few long recordings and from pieces of existing sound collections, so that pieces of the same source can appear in both training and test data.\\[6pt]
\textbf{Materials and methods:} We study the binary task of the public DroneAudioDataset (11,704 one-second clips). Its 1,332 drone clips come from 257 five-second chunks of 7 recordings, and 96.4\% of its non-drone clips are pieces of the 2,000 clips of the ESC-50 environmental sound dataset. A grouped, source-disjoint protocol (five folds, three repeats) is compared with a random split. Each clip is described by 57 hand-crafted acoustic descriptors, reduced by a hybrid minimum-redundancy maximum-relevance and ReliefF selection or by NSGA-II multi-objective search, and by a 1024-dimensional embedding from the pretrained YAMNet network. Random Forest and XGBoost learners on both views are combined by a stacking ensemble.\\[6pt]
\textbf{Results:} Under the grouped protocol, the two-view stack reached a mean accuracy of 95.29\% (standard deviation 1.62 percentage points over 15 test folds), a specificity of 98.29\% and an area under the receiver operating characteristic curve of 0.990, and it classified 98.78\% of unseen ESC-50 source clips correctly as non-drone. The random split overstated accuracy by 1.89 to 3.22 percentage points for the four models compared. Adding YAMNet embeddings to the hand-crafted view raised accuracy by 1.11 percentage points (higher in 12 of 15 folds and lower in none). Hybrid selection matched the accuracy of all 44 baseline descriptors with 16 to 18 features, and NSGA-II retained 4 to 6 features at lower accuracy. The stack processed one clip in 51.0 ms on a single CPU thread.\\[6pt]
\textbf{Conclusions:} A source-disjoint protocol removes the optimism of random clip-level splits (1.89 to 3.22 percentage points) and gives a reliable benchmark for this dataset. Fusing hand-crafted descriptors with pretrained YAMNet embeddings consistently improves detection at modest computational cost, yielding a compact, CPU-deployable detector with 16 to 18 selected features.}


\begin{document}

\section{Introduction}

Small unmanned aerial vehicles (UAVs) have become inexpensive and easy to fly, and they are used for inspection, mapping, delivery and recreation. Their growing number in low-altitude airspace, near airports and over public events has made automatic drone detection a practical need for airspace safety and traffic monitoring. Radar, radio-frequency monitoring, cameras and microphones have all been applied to the task. Acoustic sensing is attractive because the motors and rotors of a multirotor produce a stable harmonic signature~\cite{djurek2020}, microphones are cheap, and sound does not depend on lighting or a clear line of sight~\cite{dumitrescu2020}.

Acoustic drone detectors fall into two broad families. The first trains deep networks on time--frequency representations such as log-mel spectrograms or Mel-frequency cepstral coefficients (MFCCs). Djurek et al. showed that quadcopter sound carries harmonic structure that changes with the flight regime~\cite{djurek2020}, and Dumitrescu et al. developed an acoustic system for UAV detection~\cite{dumitrescu2020}. Al-Emadi et al. recorded the DroneAudioDataset and compared deep network architectures for drone detection and identification~\cite{alemadi2019}, and later enlarged the data with generative adversarial networks~\cite{alemadi2021}. Casabianca and Zhang combined several deep networks by late fusion~\cite{casabianca2021}, and Cabrera-Ponce et al. detected nearby UAVs with a microphone array carried on board a UAV~\cite{cabreraponce2020}.

The second family computes a compact set of clip-level descriptors and applies classical learners. Tian and Yu recognised UAVs in outdoor areas in real time with a Random Forest algorithm~\cite{tian2022}, Wang et al. used engineered acoustic features to detect whether a UAV carries a payload~\cite{wang2022}, and Tejera-Berengue et al. analysed how distance and environmental conditions affect machine-learning detectors~\cite{tejera2023}. The same pattern of compact descriptors, feature selection and tree-based ensembles is common in other sensing and data-analysis tasks, for example machine-learning systems built on wearable sensor data for continuous healthcare monitoring~\cite{iqbal2018} and the classification of malicious accounts on the Ethereum network with information-gain selection and gradient-boosted trees~\cite{ehsan2024}. Ensemble models have also been used to handle strongly imbalanced classes in industrial inspection, for example in wafer defect classification~\cite{batool2026}, which parallels the class-weighted training studied here. Deep models reach high accuracy but need more data and computation; descriptor-based models are cheap and interpretable but depend on the choice of features. Acoustic detection is also often combined with other modalities: deep learning on radio-frequency signals has been used for end-to-end detection and identification~\cite{alam2023}, and Frid et al. fused radio-frequency and acoustic features~\cite{frid2024}. The present study stays within the acoustic modality and fuses two views of the same signal.

Because descriptor-based detectors depend on their features, the selection step matters. Minimum-redundancy maximum-relevance (mRMR) selection adds features with high mutual information with the class and low redundancy with features already chosen~\cite{peng2005}, whereas ReliefF scores features by how well they separate near neighbours of different classes~\cite{kononenko1994,robnik2003}. Evolutionary multi-objective search, typically with NSGA-II~\cite{deb2002}, treats selection as a trade-off between error and subset size~\cite{kundu2022}, and population-based search more generally has been applied to optimisation problems in dynamic environments~\cite{iqbal2015pso}. At the same time, networks pretrained on large audio corpora such as AudioSet~\cite{gemmeke2017} now provide general-purpose embeddings at little cost. YAMNet~\cite{yamnet} is a MobileNetV1-style network~\cite{howard2017} trained with supervision on the 521 AudioSet classes, and its 1024-dimensional penultimate activations are used as transferable embeddings~\cite{yamnet,hershey2017}. Because it was not trained for drone sounds, we measure whether it adds information to a well-chosen set of hand-crafted descriptors.

A second question concerns evaluation. Public drone audio datasets are often distributed as short clips cut from a few longer recordings, and their non-drone class is often assembled from existing sound collections. When such clips are split at random, pieces of one recording or of one source sound fall on both sides of the split, and test accuracy then rewards recognising a recording rather than detecting a drone. Leakage of this kind is a known cause of over-optimistic results across machine-learning research~\cite{kapoor2023}. Shulman recently grouped the drone clips of the DroneAudioDataset by recording session and the non-drone clips by their source, and reported lower scores under grouped cross-validation than under file-level grouping~\cite{shulman2026}. The binary task of this dataset~\cite{alemadi2019} is a clear case: 96.4\% of its 10,372 non-drone clips are one-second pieces of the 2,000 clips of the ESC-50 dataset~\cite{piczak2015}, and its 1,332 drone clips come from 257 five-second chunks of only 7 recordings. ESC-50 therefore cannot act as an independent test set for models trained on this dataset, and random splits overstate performance for both classes.

This paper revisits hybrid feature selection and two-view feature fusion for acoustic drone detection under an evaluation protocol that keeps every source on one side of each split. It compares a random split with a grouped protocol on identical test clips for every model, measures how the split affects specificity on unseen ESC-50 sources, and combines the protocol with feature-selection and fusion experiments.

The main contributions of this paper are as follows:
\begin{enumerate}
	\item A leakage-aware, source-disjoint evaluation protocol for the DroneAudioDataset binary task, in which drone clips are
	grouped by recording chunk and non-drone clips by their ESC-50 source clip, with five folds and three repeats. Measured
	side by side on identical test clips, a random split overstated accuracy by 1.89 to 3.22 percentage points for the four
	models compared (Section~\ref{sec:res-protocol}, Table~\ref{tab:leakage} and Figure~\ref{fig:leakage}).
	\item A hybrid mRMR and ReliefF feature selection with an ablation against each filter and against the full descriptor
	set. It matched the accuracy of all 44 baseline descriptors (91.71\% against 91.61\%) with 16 to 18 features, and a
	selection-frequency analysis over 15 test folds shows which descriptors are chosen consistently
	(Section~\ref{sec:res-fs}, Table~\ref{tab:fs} and Figure~\ref{fig:selfreq}).
	\item A multi-objective NSGA-II search that maps the trade-off between error and the number of features, retaining 4 to 6
	features at lower accuracy for settings with tight resource limits (Section~\ref{sec:res-nsga}, Table~\ref{tab:nsga} and
	Figure~\ref{fig:nsga}).
	\item A two-view stacking ensemble of Random Forest and XGBoost learners on hand-crafted descriptors and YAMNet
	embeddings. It reached a mean accuracy of 95.29\% (standard deviation 1.62 percentage points over 15 test folds) and
	gained 1.11 percentage points over the hand-crafted stack, higher in 12 of 15 folds and lower in none, assessed with
	paired exact McNemar tests with Holm correction and bootstrap confidence intervals
	(Section~\ref{sec:res-fusion}, Tables~\ref{tab:fusion}--\ref{tab:bootstrap} and Figure~\ref{fig:fusion}).
	\item A per-class analysis on unseen ESC-50 source clips, of which 98.78\% were correctly rejected, that identifies the
	sounds most often mistaken for a drone (washing machine 20.00\%, rain 9.17\%), together with single-thread CPU latency
	of 51.0 ms per clip for the proposed stack (Section~\ref{sec:res-esc}, Tables~\ref{tab:escclasses}
	and~\ref{tab:latency}, Figures~\ref{fig:esc} and~\ref{fig:latency}).
\end{enumerate}

\section{Materials and methods}

\begin{figure}[!htbp]
\centering
\includegraphics[width=\textwidth]{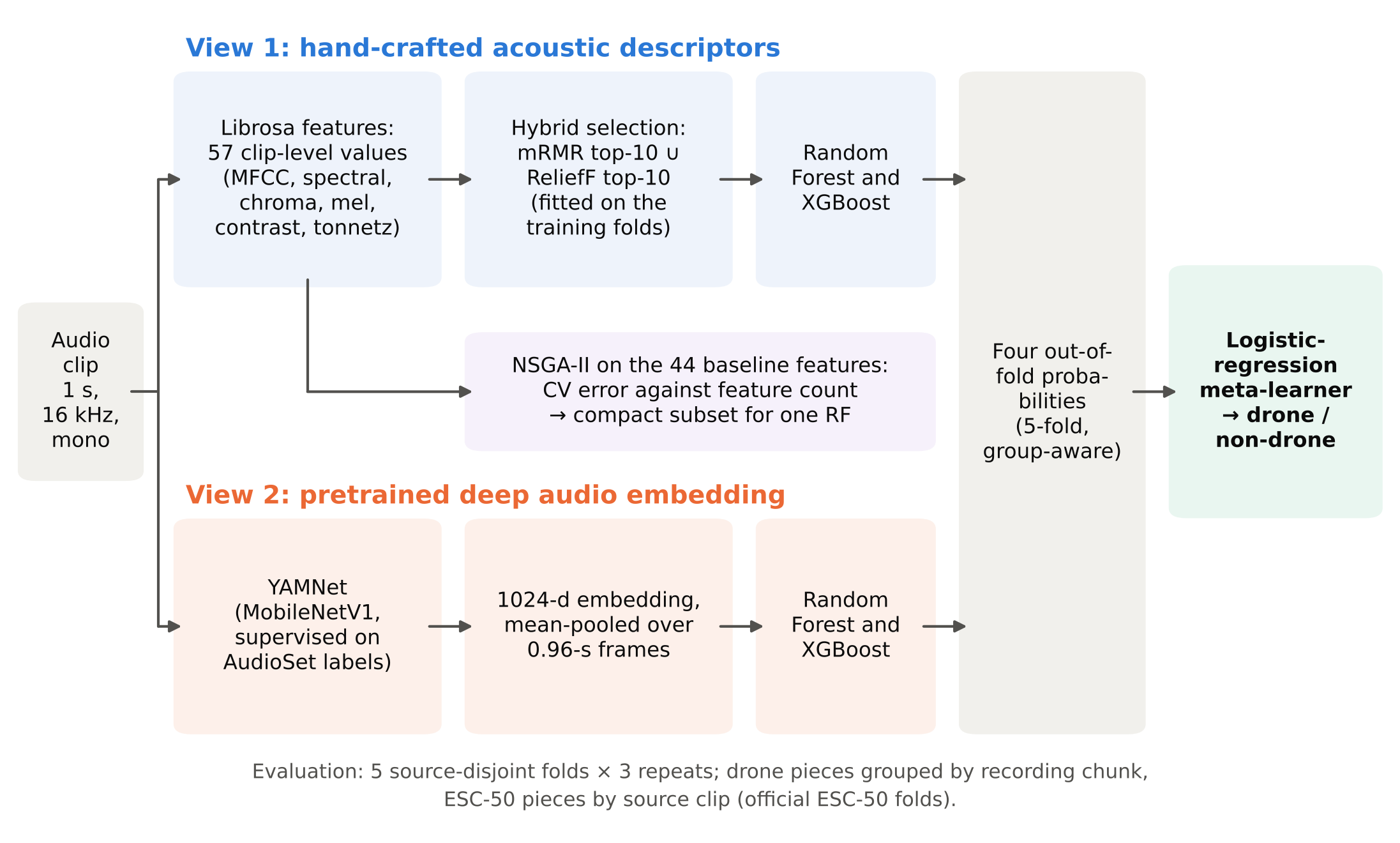}
\caption{Overview of the detection pipeline. Feature selection, base learners and the meta-learner are fitted on the training folds only.}
\label{fig:pipeline}
\end{figure}

Figure~\ref{fig:pipeline} summarises the system. Each one-second clip is described in two ways: by 57 hand-crafted descriptors, reduced by hybrid mRMR+ReliefF selection, and by a 1024-dimensional YAMNet embedding. Random Forest and XGBoost learners are trained on each view, and a logistic-regression meta-learner combines their out-of-fold probabilities. A separate NSGA-II search yields a compact subset for a single lightweight classifier. All selection and training steps are fitted on the training part of each fold only, and every model is evaluated on the same test clips, so that all comparisons in Section~\ref{sec:results} are paired.

\subsection{Dataset and leakage-aware evaluation protocol}
\label{sec:data}
We use the binary part of the DroneAudioDataset~\cite{alemadi2019,dronedataset}, which holds 1,332 drone clips and 10,372 non-drone clips of about one second, mono, at 16~kHz. According to its documentation, the drone clips are propeller noise recorded indoors from two quadcopters, labelled Bebop and Membo in the file names~\cite{dronedataset}. The file names identify the five-second chunk and the recording of every drone clip and the source of every non-drone clip: 10,001 non-drone clips are one-second pieces of the 2,000 ESC-50 clips~\cite{piczak2015}, 70 are pieces of background-noise recordings of the Speech Commands dataset~\cite{warden2018}, and 301 are silence clips. Table~\ref{tab:dataset} lists the composition. The dataset was obtained from its public repository (commit \texttt{ec6139b3}) and is used under its academic research licence.

\begin{table}[!htbp]
\caption{Composition of the binary DroneAudioDataset task (1-s clips, 16 kHz) and the groups used to keep each source on one side of every split.}
\label{tab:dataset}
\tablesize{\normalsize}
\setlength{\tabcolsep}{5pt}
\renewcommand{\arraystretch}{1.3}
\begin{tabularx}{\textwidth}{LcL}
\toprule
\textbf{Subset} & \textbf{Clips} & \textbf{Leakage groups} \\
\midrule
Drone (positive) & 1,332 & 257 five-second chunks from 7 recordings \\
\arrayrulecolor{black!25}\midrule[0.3pt]\arrayrulecolor{black}
Non-drone: ESC-50 pieces & 10,001 & 2,000 ESC-50 source clips (50 classes) \\
\arrayrulecolor{black!25}\midrule[0.3pt]\arrayrulecolor{black}
Non-drone: Speech Commands background noise & 70 & 5 noise recordings \\
\arrayrulecolor{black!25}\midrule[0.3pt]\arrayrulecolor{black}
Non-drone: silence & 301 & 1 group \\
\arrayrulecolor{black!25}\midrule[0.3pt]\arrayrulecolor{black}
Total & 11,704 &  \\
\bottomrule
\end{tabularx}
\end{table}

Two properties of this composition shape the evaluation design. The drone class rests on only 7 recordings, so neighbouring one-second pieces of one recording are acoustically very similar. The non-drone class is dominated by ESC-50, whose five-second clips each contribute about five pieces. A split that ignores these sources measures, in part, how well a model recognises a recording it has already heard.

\begin{figure}[!htbp]
\centering
\includegraphics[width=\textwidth]{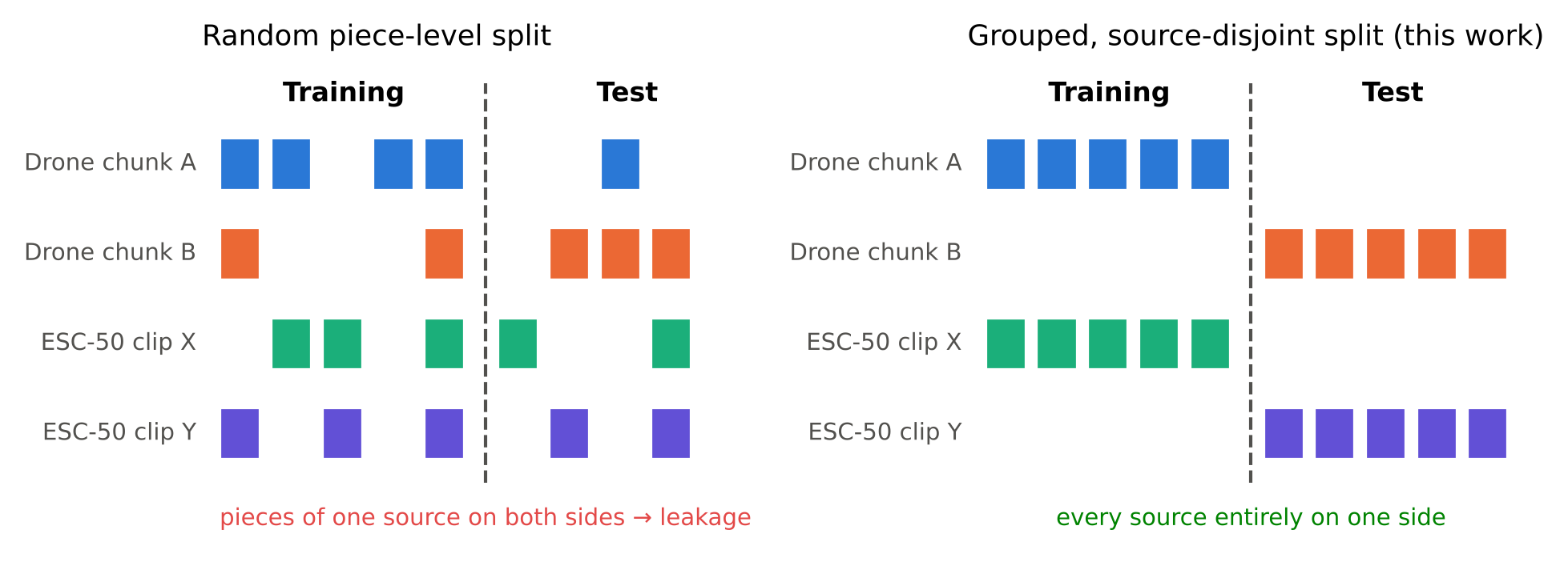}
\caption{Random piece-level splitting places pieces of one source on both sides of the split (left); the grouped protocol keeps every recording chunk and every ESC-50 source clip on one side (right).}
\label{fig:split}
\end{figure}

Two protocols are therefore compared (Figure~\ref{fig:split}). In the random piece-level split, clips are split 80/20 at random, stratified by class, three times with different seeds. Pieces of one recording chunk and of one ESC-50 clip can then appear in both training and test data, and every ESC-50 clip used for testing has contributed pieces to training. This arm reproduces the clip-level evaluation that such datasets invite, and it runs the same models with the same settings as the main protocol, so that the two protocols differ only in how clips are assigned to the test set.

The grouped, source-disjoint split is the main protocol. In it, ESC-50 pieces keep the official ESC-50 fold of their source clip, so all pieces of a clip always fall in the same fold. Drone pieces are grouped by their five-second chunk (257 groups), and the chunks of each of the 7 recordings are spread over the five folds so that the folds receive similar numbers of clips from every recording. The Speech Commands noise sources and the silence group are assigned to folds as whole groups. Each fold is the test fold once, and the assignment is repeated three times with different seeds, giving 15 test folds.

Within each training pool all drone clips are kept, and non-drone clips are sampled at a 1:1 ratio (the balanced pool) or at a 1:3 ratio with class-weighted learners (the class-weighted pool). The 1:3 ratio exposes the learners to a wider range of non-drone sounds, while the class weights keep the two classes balanced in the loss. Each test fold contains all of its drone clips and an equal number of its non-drone clips, sampled once and shared by every model; test folds held 490 to 644 clips, and training pools 2,020 to 2,174 clips (balanced) or 4,040 to 4,348 clips (class-weighted). Every ESC-50 source clip of the test fold is also scored at clip level as the mean drone probability of its pieces. Because no piece of these clips was seen in training, the share of them classified as non-drone is an unseen-source specificity, which complements the balanced test folds with a much larger and more varied set of negatives.

\subsection{Acoustic descriptors and feature selection}
\label{sec:features}
Clips are processed with librosa~\cite{mcfee2015} at 16~kHz with a 2,048-sample FFT and a hop of 512 samples. The baseline set has 44 clip-level values: the mean and standard deviation of 13 MFCCs~\cite{davis1980} (26 values); the means of spectral centroid, bandwidth, roll-off at 85\% and zero-crossing rate (4); the means of 12 chroma bins (12); and the mean and standard deviation of a 128-band log-mel spectrogram (2). The extended set adds the means of seven spectral-contrast values~\cite{jiang2002}, computed from 100~Hz so that the highest band stays below the 8~kHz Nyquist limit, and of six tonnetz dimensions~\cite{harte2006}, giving 57 features. The families describe different aspects of the same sound: MFCCs and the log-mel statistics summarise the spectral envelope, the spectral-shape measures locate the energy across frequency, and chroma, spectral contrast and tonnetz respond to tonal and harmonic structure such as the blade-passing harmonics of a rotor.

Feature selection combines two filters with different criteria. Let $X_j$ be feature $j$, $c$ the class, $S$ the set of features already selected and $I(\cdot;\cdot)$ mutual information. mRMR in its difference form adds at each step the feature that maximises
\begin{equation}
J(X_j) = I(X_j; c) - \frac{1}{|S|}\sum_{X_s \in S} I(X_j; X_s),
\end{equation}
starting from the most relevant feature~\cite{peng2005}. Mutual information is estimated with the nearest-neighbour estimators of scikit-learn~\cite{pedregosa2011} on standardised features. ReliefF updates the weight of each feature from the $k$ nearest hits $H_i$ and misses $M_i$ of each sampled instance $R$,
\begin{equation}
W(X_j) \leftarrow W(X_j) - \sum_{i=1}^{k}\frac{\mathrm{diff}(X_j,R,H_i)}{mk} + \sum_{i=1}^{k}\frac{\mathrm{diff}(X_j,R,M_i)}{mk},
\end{equation}
where $\mathrm{diff}$ is the range-normalised absolute difference and $m$ the number of sampled instances~\cite{kononenko1994,robnik2003}. We use $k=100$ on min--max-scaled features, as implemented in scikit-rebate~\cite{urbanowicz2018}. Both estimates use a stratified sample of at most 3,000 training clips. mRMR favours features that are informative and not redundant with each other, whereas ReliefF rewards features that separate close examples of the two classes and can keep features that are useful only in combination. The hybrid subset is the union of the ten top-ranked features of each method, which gave 16 to 18 features per fold.

Filter selection fixes the subset size in advance. To examine how few descriptors are sufficient, NSGA-II~\cite{deb2002} searches binary masks $z \in \{0,1\}^{44}$ over the baseline descriptors and minimises
\begin{equation}
f_1(z) = 1 - \mathrm{Acc}_{\mathrm{CV}}(z), \qquad f_2(z) = \textstyle\sum_{j} z_j,
\end{equation}
where $\mathrm{Acc}_{\mathrm{CV}}$ is the grouped three-fold cross-validation accuracy of a 100-tree Random Forest inside the training pool. The search uses a population of 40, 40 generations, two-point crossover and bit-flip mutation (pymoo~\cite{blank2020}). The operating point was fixed before testing as the smallest subset on the final front whose inner accuracy lies within two percentage points of the best one. Because of its cost, NSGA-II was run on the five folds of the first repeat.

\subsection{Pretrained embedding and stacking ensemble}
\label{sec:models}
YAMNet~\cite{yamnet} frames the 16~kHz waveform into 0.96-s windows with a 0.48-s hop and returns a 1024-dimensional embedding per window; clips shorter than one window are zero-padded. The window embeddings are averaged into one vector per clip. YAMNet is used as a frozen feature extractor without fine-tuning, so it provides a second view of each clip that was learned from a large general-purpose audio corpus rather than designed for this task.

Baselines on the hybrid subset are logistic regression on standardised inputs, a Random Forest tuned by grouped grid search (100 or 200 trees, maximum depth unlimited, 10 or 20, minimum split size 2 or 5), XGBoost (300 trees, learning rate 0.05, depth 6) and a one-dimensional convolutional network (two convolutional layers with 32 and 64 filters, global average pooling and two dense layers, Adam, early stopping on 10\% of the training data). The stacking ensemble uses four base learners, a Random Forest of 200 trees and XGBoost on the hand-crafted view and the same two learners on YAMNet embeddings. In the class-weighted pool the Random Forests use balanced class weights and XGBoost weights the positive class by the class ratio. Out-of-fold probabilities $p_1,\dots,p_4$ from five-fold grouped cross-validation inside the training pool train a logistic-regression meta-learner,
\begin{equation}
\hat{P}(\mathrm{drone}\mid x) = \sigma\Big(\beta_0 + \sum_{m=1}^{4} \beta_m\, p_m(x)\Big),
\end{equation}
after which the base learners are refitted on the whole training pool~\cite{wolpert1992}. Because the meta-learner sees only probabilities for clips that its base learners did not train on, it weights each learner by how well it generalises rather than by how well it fits the training pool. A clip is labelled as drone when its probability is at least 0.5. The proposed model, fixed before any result was seen, is the two-view stack trained on the class-weighted pool; the hand-crafted view of this stack uses the hybrid subset selected from all 57 descriptors (17 to 18 features). The balanced-pool stacks in Table~\ref{tab:ratio} use the hybrid subset of either the 44 baseline features or the 57 extended features.

\subsection{Performance metrics and statistical analysis}
Drone is the positive class. With TP, TN, FP and FN the numbers of true positives, true negatives, false positives and false negatives, recall is TP/(TP+FN), specificity TN/(TN+FP), precision TP/(TP+FP), and the F1-score is the harmonic mean of precision and recall. We report these values, accuracy and the area under the ROC curve (AUC) as mean $\pm$ standard deviation over the 15 grouped test folds; because test folds are balanced, accuracy equals balanced accuracy. ESC-50 unseen-source specificity is pooled over the five folds of each repeat and summarised over the three repeats.

Fold accuracies are not independent, because the training pools of different folds overlap. The proposed stack is therefore compared with each alternative on identical clips by the exact McNemar test, which uses only the clips on which the two models disagree, applied to the predictions of each repeat with its folds pooled~\cite{dietterich1998}. Holm correction controls the family-wise error rate over the five comparisons within each repeat~\cite{holm1979}, and per-fold paired differences with won, tied and lost counts are reported alongside. Pooled accuracies carry 95\% bootstrap confidence intervals from 2,000 resamples~\cite{efron1993}. Latency is measured on one CPU thread as the median over 100 clips and three repetitions.

\subsection{Implementation}
The pipeline is written in Python with librosa, scikit-learn, XGBoost, pymoo, scikit-rebate and TensorFlow~\cite{abadi2016}. Experiments ran on a workstation with two AMD EPYC 7513 processors. All 18 runs (15 grouped folds and 3 random splits) were produced by one version of the code, and every table and figure in Section~\ref{sec:results} was generated directly from the saved per-fold predictions. Latency is the median over 100 clips repeated three times on one CPU thread, including feature extraction and, for the two-view stack, YAMNet; model size is the serialised classifier.

\section{Results}
\label{sec:results}

This section first measures how the evaluation protocol changes the reported performance. All later experiments use the grouped protocol: single classifiers and feature selection (Section~\ref{sec:res-fs}), two-view fusion and the training class ratio (Section~\ref{sec:res-fusion}), and the errors and computational cost of the proposed stack (Section~\ref{sec:res-esc}).

\subsection{Effect of the evaluation protocol}
\label{sec:res-protocol}
\begin{figure}[!htbp]
\centering
\includegraphics[width=\textwidth]{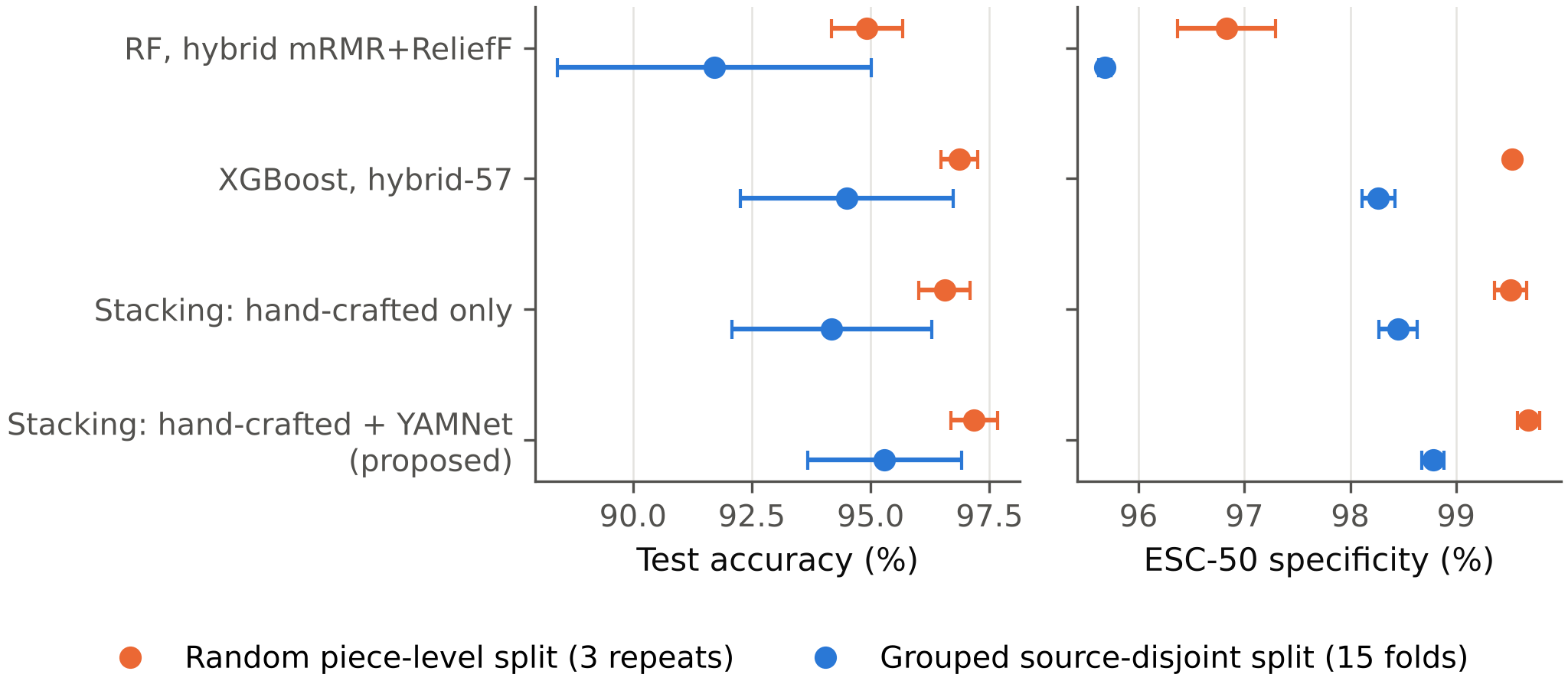}
\caption{Accuracy and ESC-50 specificity under the random and grouped protocols (mean $\pm$ SD).}
\label{fig:leakage}
\end{figure}

Table~\ref{tab:leakage} and Figure~\ref{fig:leakage} compare the two protocols for four representative models. The random split gave higher accuracy for every model, by 1.89 to 3.22 percentage points, because its test clips share recording chunks and ESC-50 sources with the training data. For the proposed stack, accuracy fell from 97.18\% under the random split to 95.29\% under the grouped protocol.

\begin{table}[!htbp]
\caption{Effect of the evaluation protocol.}
\label{tab:leakage}
\tablesize{\small}
\setlength{\tabcolsep}{3pt}
\renewcommand{\arraystretch}{1.3}
\begin{tabularx}{\textwidth}{Lccccc}
\toprule
 & \multicolumn{3}{c}{\textbf{Accuracy (\%)}} & \multicolumn{2}{c}{\textbf{ESC-50 specificity (\%)}} \\
\cmidrule(lr){2-4}\cmidrule(lr){5-6}
\textbf{Model} & Random & Grouped & $\Delta$ & Random & Grouped \\
\midrule
RF, hybrid mRMR+\allowbreak{}ReliefF & 94.92 $\pm$ 0.75 & 91.71 $\pm$ 3.30 & 3.22 & 96.83 $\pm$ 0.46 & 95.68 $\pm$ 0.06 \\
\arrayrulecolor{black!25}\midrule[0.3pt]\arrayrulecolor{black}
XGBoost, hybrid of 57 features & 96.87 $\pm$ 0.39 & 94.50 $\pm$ 2.24 & 2.37 & 99.53 $\pm$ 0.03 & 98.27 $\pm$ 0.15 \\
\arrayrulecolor{black!25}\midrule[0.3pt]\arrayrulecolor{black}
Stacking: hand-crafted only & 96.55 $\pm$ 0.54 & 94.18 $\pm$ 2.10 & 2.37 & 99.52 $\pm$ 0.15 & 98.45 $\pm$ 0.18 \\
\arrayrulecolor{black!25}\midrule[0.3pt]\arrayrulecolor{black}
Stacking: hand-crafted + YAMNet (proposed) & 97.18 $\pm$ 0.50 & 95.29 $\pm$ 1.62 & 1.89 & 99.68 $\pm$ 0.10 & 98.78 $\pm$ 0.10 \\
\bottomrule
\end{tabularx}
\par\smallskip{\footnotesize Random: 80/20 split of 1-s clips (3 repeats). Grouped: source-disjoint protocol (15 folds). Mean $\pm$ SD; $\Delta$ = random minus grouped accuracy in percentage points.}
\end{table}

The ranking of the four models was the same under both protocols, so the random split did not reorder them; it moved all of them upward, by different amounts. The shift was largest for the Random Forest on the hybrid subset (3.22 points) and smallest for the proposed stack (1.89 points). Even the smallest shift is larger than every difference between the three strongest models under the grouped protocol, the largest of which is 1.11 points. The random split also reported standard deviations three to six times smaller than the grouped protocol, so it overstates both the level and the stability of performance.

ESC-50 specificity followed the same pattern. Under the random split the three strongest models classified 99.52\% to 99.68\% of ESC-50 clips as non-drone, but these clips had contributed other pieces to training. When the clips were truly unseen, specificity fell to 98.78\% for the proposed stack (Figure~\ref{fig:leakage}, right). The grouped values are the ones used in the rest of the paper.

\subsection{Single classifiers and feature selection}
\label{sec:res-fs}
\label{sec:res-nsga}
Table~\ref{tab:baselines} lists the single classifiers on the hybrid subset under the grouped protocol. XGBoost was the most accurate (94.10\%), followed by the tuned Random Forest (91.71\%), logistic regression (91.12\%) and the one-dimensional CNN (88.96\%). The Random Forest had high specificity but the lowest recall of the four classifiers (88.57~$\pm$~7.05\%), with a large spread across folds. Logistic regression and the CNN showed the opposite profile, with recall above specificity (92.35\% against 89.90\% for logistic regression), and they rejected fewer unseen ESC-50 sources (91.12\% and 89.88\%) than the two tree ensembles (95.68\% and 95.50\%). On a few hand-crafted inputs, the tree ensembles were thus the more reliable single learners, and they are also the base learners of the stacking ensemble.

\begin{table}[!htbp]
\caption{Classifiers on the hybrid mRMR+ReliefF feature subset (balanced pool).}
\label{tab:baselines}
\tablesize{\small}
\setlength{\tabcolsep}{3pt}
\renewcommand{\arraystretch}{1.3}
\begin{tabularx}{\textwidth}{Lccccccc}
\toprule
\textbf{Model} & \textbf{Acc.} & \textbf{Prec.} & \textbf{Rec.} & \textbf{F1} & \textbf{Spec.} & \textbf{AUC} & \textbf{ESC-50 spec.} \\
\midrule
Logistic regression, hybrid & 91.12 $\pm$ 2.13 & 90.25 $\pm$ 2.95 & 92.35 $\pm$ 3.48 & 91.23 $\pm$ 2.11 & 89.90 $\pm$ 3.61 & 0.965 $\pm$ 0.016 & 91.12 $\pm$ 0.29 \\
\arrayrulecolor{black!25}\midrule[0.3pt]\arrayrulecolor{black}
1-D CNN, hybrid & 88.96 $\pm$ 3.19 & 88.50 $\pm$ 5.43 & 90.13 $\pm$ 5.14 & 89.10 $\pm$ 2.98 & 87.79 $\pm$ 6.69 & 0.957 $\pm$ 0.025 & 89.88 $\pm$ 1.83 \\
\arrayrulecolor{black!25}\midrule[0.3pt]\arrayrulecolor{black}
RF, hybrid mRMR+\allowbreak{}ReliefF & 91.71 $\pm$ 3.30 & 94.57 $\pm$ 1.96 & 88.57 $\pm$ 7.05 & 91.31 $\pm$ 3.86 & 94.85 $\pm$ 2.07 & 0.981 $\pm$ 0.010 & 95.68 $\pm$ 0.06 \\
\arrayrulecolor{black!25}\midrule[0.3pt]\arrayrulecolor{black}
XGBoost, hybrid & 94.10 $\pm$ 1.78 & 94.80 $\pm$ 1.58 & 93.36 $\pm$ 3.84 & 94.03 $\pm$ 1.92 & 94.84 $\pm$ 1.68 & 0.985 $\pm$ 0.009 & 95.50 $\pm$ 0.10 \\
\bottomrule
\end{tabularx}
\par\smallskip{\footnotesize Mean $\pm$ SD over 15 grouped test folds, in percent (AUC as a fraction); ESC-50 spec.: unseen-source specificity, mean $\pm$ SD over 3 repeats.}
\end{table}

Table~\ref{tab:fs} isolates the selection step with the tuned Random Forest. The hybrid subset reached 91.71~$\pm$~3.30\% accuracy, against 91.61~$\pm$~2.47\% with all 44 descriptors, 91.33\% with the mRMR top ten and 90.69\% with the ReliefF top ten. The hybrid subset matched the full set within one standard deviation while using fewer than half of the descriptors. Each single filter fell slightly below the hybrid in accuracy and in ESC-50 specificity (94.42\% and 94.88\% against 95.68\%), which supports combining the two criteria; the full set kept a slightly higher ESC-50 specificity (96.17\%).

\begin{table}[!htbp]
\caption{Feature-selection ablation with a tuned Random Forest (balanced pool).}
\label{tab:fs}
\tablesize{\small}
\setlength{\tabcolsep}{3pt}
\renewcommand{\arraystretch}{1.3}
\begin{tabularx}{\textwidth}{Lccccccc}
\toprule
\textbf{Model} & \textbf{Acc.} & \textbf{Prec.} & \textbf{Rec.} & \textbf{F1} & \textbf{Spec.} & \textbf{AUC} & \textbf{ESC-50 spec.} \\
\midrule
RF, all 44 baseline features & 91.61 $\pm$ 2.47 & 94.82 $\pm$ 2.57 & 88.12 $\pm$ 5.06 & 91.24 $\pm$ 2.81 & 95.09 $\pm$ 2.63 & 0.981 $\pm$ 0.010 & 96.17 $\pm$ 0.39 \\
\arrayrulecolor{black!25}\midrule[0.3pt]\arrayrulecolor{black}
RF, mRMR top-10 & 91.33 $\pm$ 2.54 & 92.39 $\pm$ 2.69 & 90.15 $\pm$ 4.55 & 91.19 $\pm$ 2.75 & 92.50 $\pm$ 2.92 & 0.973 $\pm$ 0.014 & 94.42 $\pm$ 0.46 \\
\arrayrulecolor{black!25}\midrule[0.3pt]\arrayrulecolor{black}
RF, ReliefF top-10 & 90.69 $\pm$ 3.65 & 93.42 $\pm$ 2.10 & 87.60 $\pm$ 7.51 & 90.25 $\pm$ 4.22 & 93.78 $\pm$ 2.20 & 0.973 $\pm$ 0.014 & 94.88 $\pm$ 0.23 \\
\arrayrulecolor{black!25}\midrule[0.3pt]\arrayrulecolor{black}
RF, hybrid mRMR+\allowbreak{}ReliefF & 91.71 $\pm$ 3.30 & 94.57 $\pm$ 1.96 & 88.57 $\pm$ 7.05 & 91.31 $\pm$ 3.86 & 94.85 $\pm$ 2.07 & 0.981 $\pm$ 0.010 & 95.68 $\pm$ 0.06 \\
\bottomrule
\end{tabularx}
\par\smallskip{\footnotesize Mean $\pm$ SD over 15 grouped test folds, in percent (AUC as a fraction); ESC-50 spec.: unseen-source specificity, mean $\pm$ SD over 3 repeats.}
\end{table}

\begin{figure}[!htbp]
\centering
\includegraphics[width=0.8\textwidth]{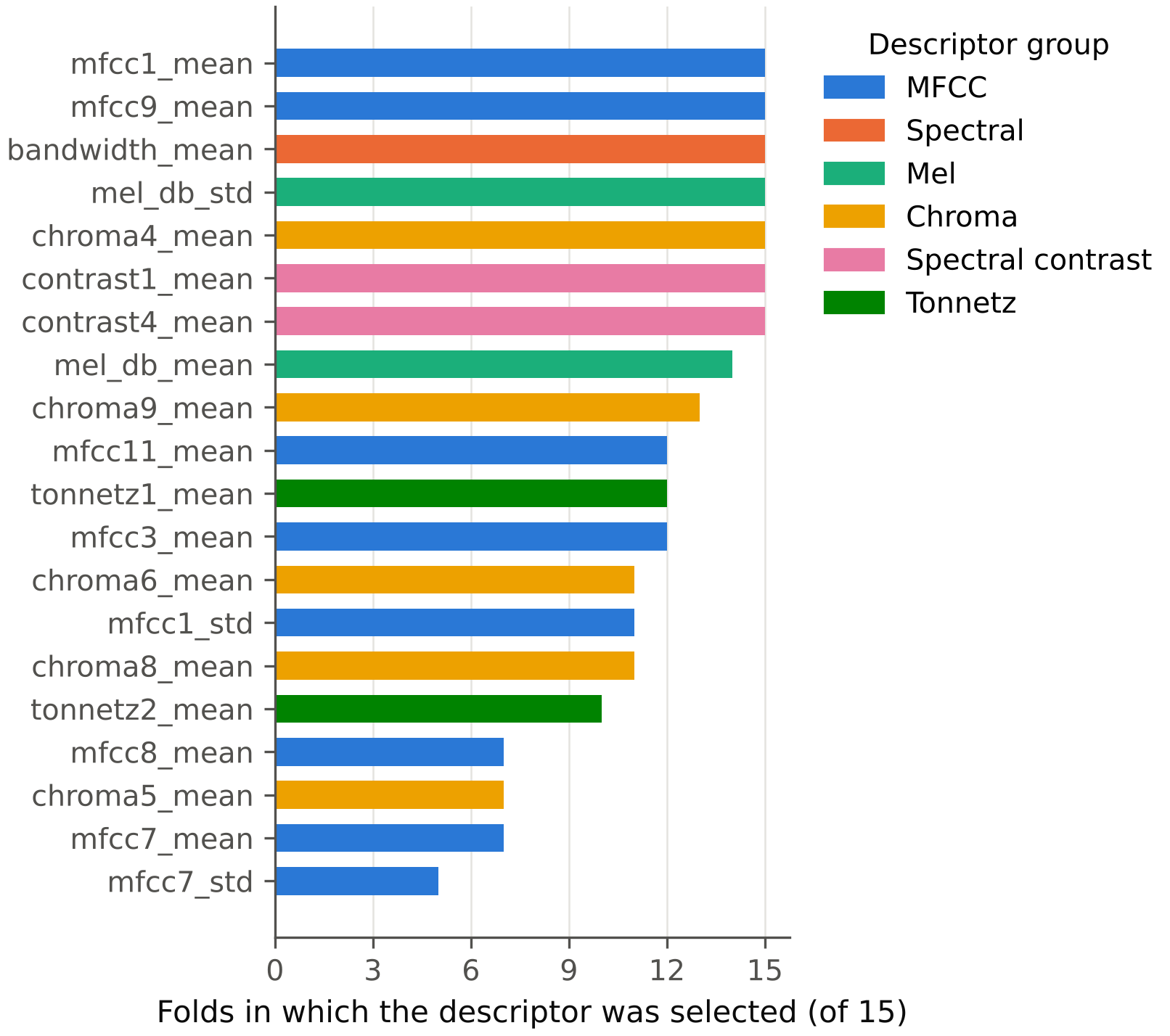}
\caption{Number of grouped test folds (of 15) in which each descriptor entered the hybrid mRMR+ReliefF subset (class-weighted pool).}
\label{fig:selfreq}
\end{figure}

Figure~\ref{fig:selfreq} shows how often each descriptor entered the hybrid subset of the class-weighted pool. The first and ninth MFCC means, spectral bandwidth, the spread of the log-mel spectrogram, one chroma bin and two spectral-contrast bands were selected in all 15 folds. MFCC statistics formed the largest group of selected descriptors, yet every descriptor family contributed at least one descriptor that was selected in most folds. The selection was stable across folds, and the information it kept is spread over several families rather than concentrated in the cepstral coefficients.

\begin{figure}[!htbp]
\centering
\includegraphics[width=0.65\textwidth]{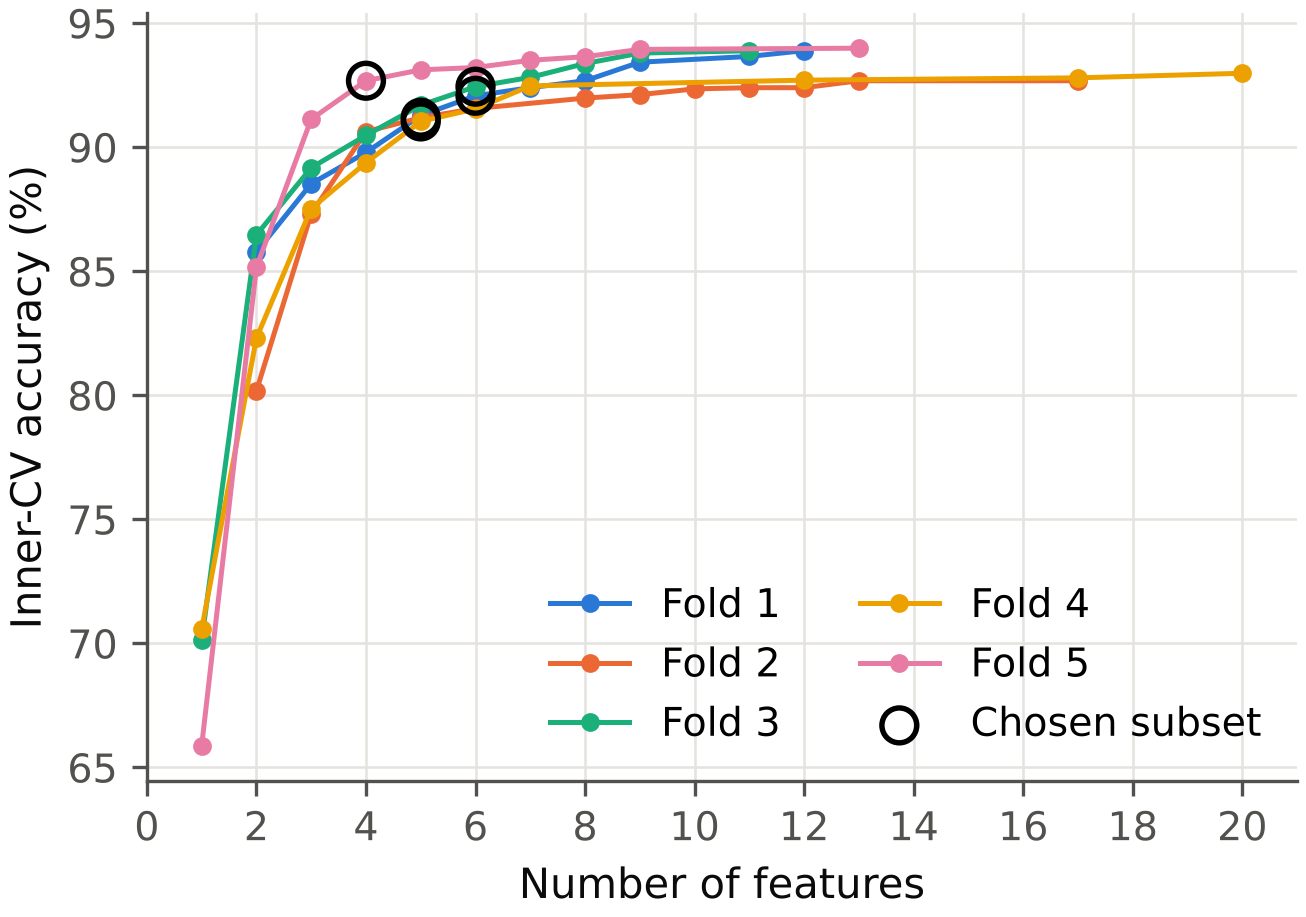}
\caption{NSGA-II Pareto fronts (inner cross-validation accuracy against subset size) for the five grouped folds of the first repeat; circles mark the pre-specified operating points.}
\label{fig:nsga}
\end{figure}

The NSGA-II search asks a different question: how few descriptors are enough? The Pareto fronts in Figure~\ref{fig:nsga} rise steeply up to about five features and flatten beyond, and the fronts of the five folds lie close together. The pre-specified rule chose 4 to 6 features per fold (Table~\ref{tab:nsga}), mostly MFCC statistics. These compact subsets reached 86.54 to 91.54\% test accuracy (mean 88.14\%), against 87.38 to 94.62\% (mean 91.37\%) for the hybrid subset on the same folds. NSGA-II thus gives up about three percentage points of accuracy for a descriptor set roughly one third the size of the hybrid subset.

\begin{table}[!htbp]
\caption{NSGA-II feature selection on the 44 baseline features (five grouped folds of the first repeat).}
\label{tab:nsga}
\tablesize{\small}
\setlength{\tabcolsep}{3pt}
\renewcommand{\arraystretch}{1.3}
\begin{tabularx}{\textwidth}{ZZZZZZZZ}
\toprule
\textbf{Fold} & \textbf{Front size} & \textbf{Chosen $k$} & \textbf{Inner-CV acc.} & \textbf{Test acc.} & \textbf{Test AUC} & \textbf{Hybrid $k$} & \textbf{Hybrid test acc.} \\
\midrule
1 & 10 & 6 & 92.07 & 88.46 & 0.953 & 17 & 92.69 \\
\arrayrulecolor{black!25}\midrule[0.3pt]\arrayrulecolor{black}
2 & 12 & 5 & 91.19 & 91.54 & 0.972 & 16 & 94.62 \\
\arrayrulecolor{black!25}\midrule[0.3pt]\arrayrulecolor{black}
3 & 10 & 6 & 92.44 & 86.97 & 0.959 & 16 & 92.15 \\
\arrayrulecolor{black!25}\midrule[0.3pt]\arrayrulecolor{black}
4 & 10 & 5 & 91.04 & 87.20 & 0.951 & 18 & 90.00 \\
\arrayrulecolor{black!25}\midrule[0.3pt]\arrayrulecolor{black}
5 & 10 & 4 & 92.68 & 86.54 & 0.968 & 17 & 87.38 \\
\bottomrule
\end{tabularx}
\par\smallskip{\footnotesize Accuracies in percent; hybrid: mRMR+ReliefF subset of the same fold.}
\end{table}

\subsection{Two-view fusion and training class ratio}
\label{sec:res-fusion}
\begin{figure}[!htbp]
\centering
\includegraphics[width=\textwidth]{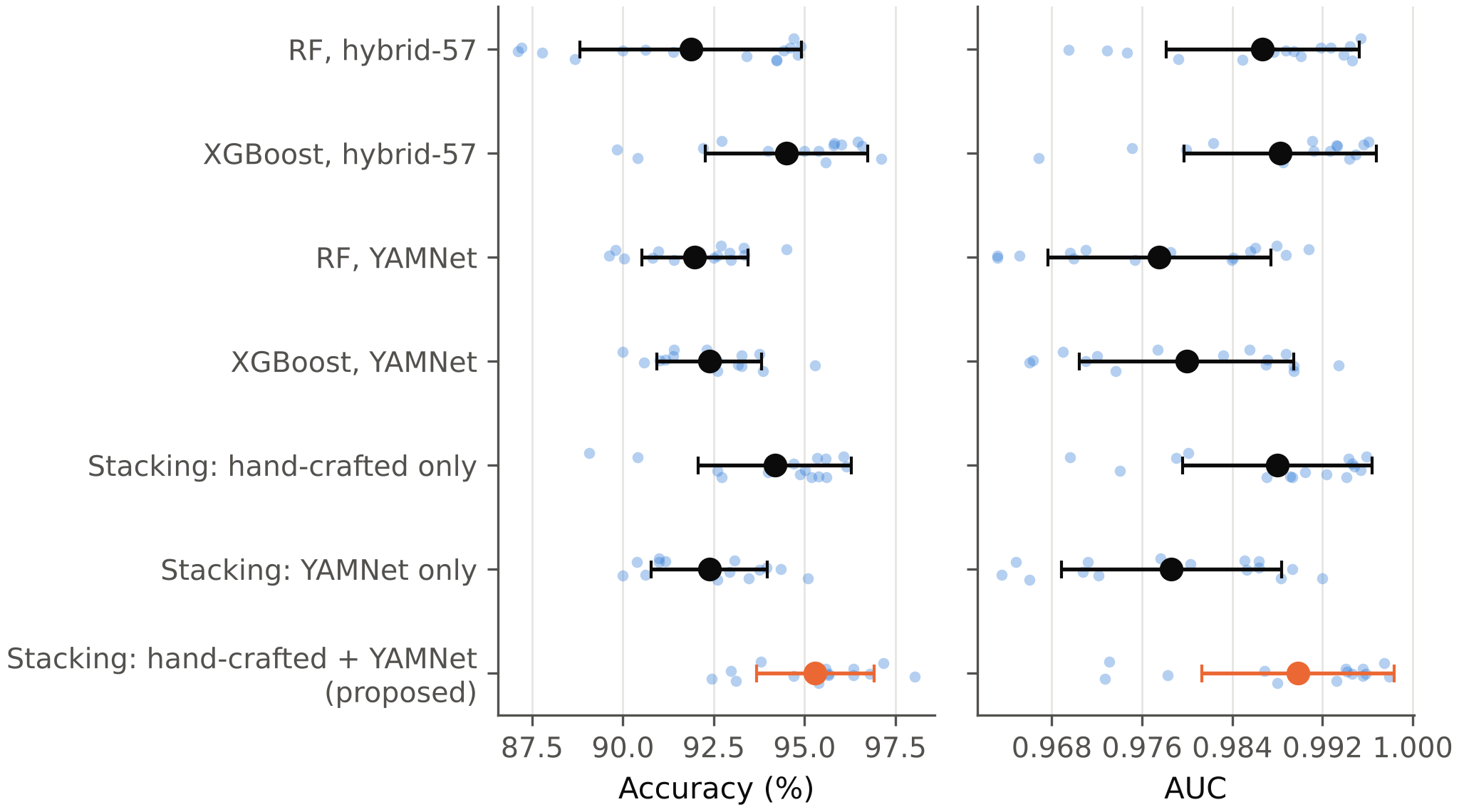}
\caption{Fusion ablation with the class-weighted pool: accuracy and AUC per grouped test fold (dots) with mean $\pm$ SD (markers).}
\label{fig:fusion}
\end{figure}

Table~\ref{tab:fusion} and Figure~\ref{fig:fusion} report the fusion ablation for the class-weighted pool. Single learners on YAMNet embeddings (91.98\% and 92.38\%) were less accurate than XGBoost on the hand-crafted subset (94.50\%). Stacking within one view gave 94.18\% (hand-crafted) and 92.38\% (YAMNet), and the two-view stack reached 95.29~$\pm$~1.62\% with 98.29\% specificity and an AUC of 0.990.

\begin{table}[!htbp]
\caption{Fusion ablation (class-weighted pool): single learners, single-view stacks and the two-view stack.}
\label{tab:fusion}
\tablesize{\small}
\setlength{\tabcolsep}{3pt}
\renewcommand{\arraystretch}{1.3}
\begin{tabularx}{\textwidth}{Lccccccc}
\toprule
\textbf{Model} & \textbf{Acc.} & \textbf{Prec.} & \textbf{Rec.} & \textbf{F1} & \textbf{Spec.} & \textbf{AUC} & \textbf{ESC-50 spec.} \\
\midrule
RF, hybrid of 57 features & 91.87 $\pm$ 3.05 & 97.61 $\pm$ 1.02 & 85.86 $\pm$ 6.31 & 91.24 $\pm$ 3.61 & 97.88 $\pm$ 0.95 & 0.987 $\pm$ 0.009 & 98.80 $\pm$ 0.13 \\
\arrayrulecolor{black!25}\midrule[0.3pt]\arrayrulecolor{black}
XGBoost, hybrid of 57 features & 94.50 $\pm$ 2.24 & 97.51 $\pm$ 0.89 & 91.33 $\pm$ 4.46 & 94.26 $\pm$ 2.50 & 97.66 $\pm$ 0.87 & 0.988 $\pm$ 0.009 & 98.27 $\pm$ 0.15 \\
\arrayrulecolor{black!25}\midrule[0.3pt]\arrayrulecolor{black}
RF, YAMNet & 91.98 $\pm$ 1.46 & 96.73 $\pm$ 1.15 & 86.90 $\pm$ 2.77 & 91.53 $\pm$ 1.64 & 97.05 $\pm$ 1.06 & 0.978 $\pm$ 0.010 & 98.25 $\pm$ 0.09 \\
\arrayrulecolor{black!25}\midrule[0.3pt]\arrayrulecolor{black}
XGBoost, YAMNet & 92.38 $\pm$ 1.44 & 96.90 $\pm$ 0.99 & 87.57 $\pm$ 2.82 & 91.97 $\pm$ 1.61 & 97.19 $\pm$ 0.92 & 0.980 $\pm$ 0.010 & 98.58 $\pm$ 0.10 \\
\arrayrulecolor{black!25}\midrule[0.3pt]\arrayrulecolor{black}
Stacking: hand-crafted only & 94.18 $\pm$ 2.10 & 97.68 $\pm$ 0.95 & 90.52 $\pm$ 4.35 & 93.91 $\pm$ 2.39 & 97.84 $\pm$ 0.94 & 0.988 $\pm$ 0.008 & 98.45 $\pm$ 0.18 \\
\arrayrulecolor{black!25}\midrule[0.3pt]\arrayrulecolor{black}
Stacking: YAMNet only & 92.38 $\pm$ 1.60 & 97.19 $\pm$ 1.03 & 87.29 $\pm$ 3.05 & 91.95 $\pm$ 1.79 & 97.47 $\pm$ 0.94 & 0.979 $\pm$ 0.010 & 98.50 $\pm$ 0.22 \\
\arrayrulecolor{black!25}\midrule[0.3pt]\arrayrulecolor{black}
Stacking: hand-crafted + YAMNet (proposed) & 95.29 $\pm$ 1.62 & 98.19 $\pm$ 0.92 & 92.30 $\pm$ 3.14 & 95.12 $\pm$ 1.76 & 98.29 $\pm$ 0.89 & 0.990 $\pm$ 0.009 & 98.78 $\pm$ 0.10 \\
\bottomrule
\end{tabularx}
\par\smallskip{\footnotesize Mean $\pm$ SD over 15 grouped test folds, in percent (AUC as a fraction); ESC-50 spec.: unseen-source specificity, mean $\pm$ SD over 3 repeats.}
\end{table}

Most of the gain came from recall. The two-view stack detected 92.30\% of drone clips, against 90.52\% for the hand-crafted stack and 87.29\% for the YAMNet stack, while its specificity changed little (98.29\% against 97.84\%). Its fold-to-fold spread was also smaller than that of the hand-crafted stack (standard deviation 1.62 against 2.10 percentage points), as the per-fold dots in Figure~\ref{fig:fusion} show.

\begin{figure}[!htbp]
\centering
\includegraphics[width=0.6\textwidth]{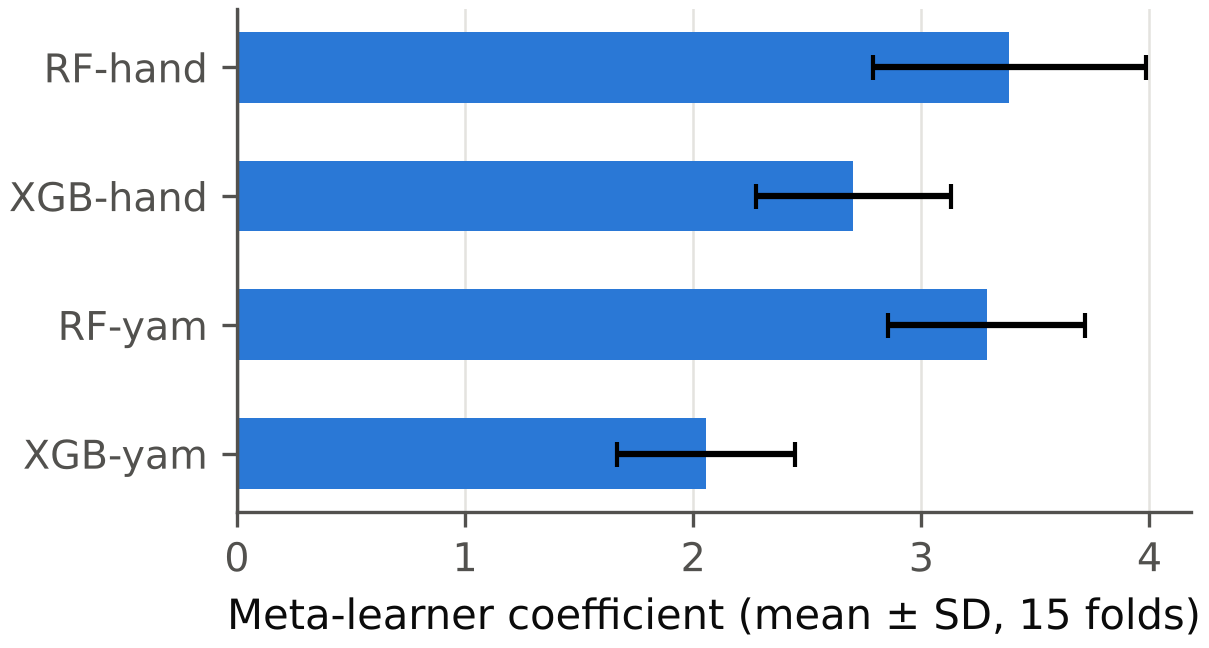}
\caption{Logistic-regression meta-learner weights of the proposed stack (mean $\pm$ SD over 15 folds).}
\label{fig:coef}
\end{figure}

The meta-learner gave similar weights to the hand-crafted and YAMNet Random Forests (3.39 and 3.29) and smaller weights to the two XGBoost learners (2.70 and 2.06), so both views contribute to the decision (Figure~\ref{fig:coef}). The YAMNet Random Forest received almost the same weight as its hand-crafted counterpart although it was the weaker learner on its own, which indicates that its errors overlap only partly with those of the hand-crafted view.

Tables~\ref{tab:delta} and~\ref{tab:mcnemar} give the paired comparisons. Against the hand-crafted stack, the two-view stack gained 1.11 percentage points on average and won 12 folds, tied 3 and lost 0; the McNemar test remained significant after Holm correction in two of the three repeats (largest adjusted $p$ = 0.074). Against XGBoost on the same hand-crafted subset the gain was 0.80 percentage points (11 won, 1 tied, 3 lost; significant in two of the three repeats). Against the YAMNet-only stack (2.91 percentage points) and the Random Forest baseline (3.59 percentage points) the differences were significant in every repeat. The gain from fusion is therefore small but consistent across folds.

\begin{table}[!htbp]
\caption{Per-fold paired accuracy differences between the proposed stack and each alternative over the 15 grouped test folds (computed from unrounded per-fold accuracies).}
\label{tab:delta}
\tablesize{\normalsize}
\setlength{\tabcolsep}{5pt}
\renewcommand{\arraystretch}{1.3}
\begin{tabularx}{\textwidth}{Lcc}
\toprule
\textbf{Alternative} & \textbf{$\Delta$ acc. (pp), mean $\pm$ SD} & \textbf{Folds won / tied / lost} \\
\midrule
Stacking: hand-crafted only & 1.11 $\pm$ 1.08 & 12 / 3 / 0 \\
\arrayrulecolor{black!25}\midrule[0.3pt]\arrayrulecolor{black}
XGBoost, hybrid of 57 features & 0.80 $\pm$ 1.08 & 11 / 1 / 3 \\
\arrayrulecolor{black!25}\midrule[0.3pt]\arrayrulecolor{black}
Stacking: YAMNet only & 2.91 $\pm$ 1.74 & 14 / 0 / 1 \\
\arrayrulecolor{black!25}\midrule[0.3pt]\arrayrulecolor{black}
RF, hybrid mRMR+\allowbreak{}ReliefF & 3.59 $\pm$ 3.53 & 13 / 0 / 2 \\
\arrayrulecolor{black!25}\midrule[0.3pt]\arrayrulecolor{black}
XGBoost, hybrid & 1.19 $\pm$ 1.55 & 12 / 0 / 3 \\
\bottomrule
\end{tabularx}
\end{table}
\begin{table}[!htbp]
\caption{Exact McNemar tests of the proposed stack against each alternative on identical test clips, with the five folds of each repeat pooled (2,664 clips per repeat). b/c: clips classified correctly only by the proposed stack / only by the alternative; Holm: $p$ corrected over the five comparisons in a repeat.}
\label{tab:mcnemar}
\tablesize{\small}
\setlength{\tabcolsep}{3pt}
\renewcommand{\arraystretch}{1.3}
\begin{tabularx}{\textwidth}{Lccccccccc}
\toprule
 & \multicolumn{3}{c}{\textbf{Repeat 1}} & \multicolumn{3}{c}{\textbf{Repeat 2}} & \multicolumn{3}{c}{\textbf{Repeat 3}} \\
\cmidrule(lr){2-4}\cmidrule(lr){5-7}\cmidrule(lr){8-10}
\textbf{Alternative} & b/c & $p$ & Holm & b/c & $p$ & Holm & b/c & $p$ & Holm \\
\midrule
Stacking: hand-crafted only & 55 / 33 & 0.025 & 0.074 & 55 / 27 & 0.003 & 0.008 & 57 / 19 & $<$ 0.001 & $<$ 0.001 \\
\arrayrulecolor{black!25}\midrule[0.3pt]\arrayrulecolor{black}
XGBoost, hybrid of 57 features & 53 / 40 & 0.213 & 0.213 & 52 / 28 & 0.010 & 0.012 & 48 / 25 & 0.010 & 0.010 \\
\arrayrulecolor{black!25}\midrule[0.3pt]\arrayrulecolor{black}
Stacking: YAMNet only & 98 / 23 & $<$ 0.001 & $<$ 0.001 & 114 / 32 & $<$ 0.001 & $<$ 0.001 & 108 / 34 & $<$ 0.001 & $<$ 0.001 \\
\arrayrulecolor{black!25}\midrule[0.3pt]\arrayrulecolor{black}
RF, hybrid mRMR+\allowbreak{}ReliefF & 151 / 45 & $<$ 0.001 & $<$ 0.001 & 143 / 47 & $<$ 0.001 & $<$ 0.001 & 145 / 39 & $<$ 0.001 & $<$ 0.001 \\
\arrayrulecolor{black!25}\midrule[0.3pt]\arrayrulecolor{black}
XGBoost, hybrid & 76 / 55 & 0.080 & 0.160 & 95 / 60 & 0.006 & 0.012 & 95 / 48 & $<$ 0.001 & $<$ 0.001 \\
\bottomrule
\end{tabularx}
\end{table}

Table~\ref{tab:bootstrap} gives the pooled accuracy of each repeat with its 95\% bootstrap confidence interval. The intervals of the two-view stack and the hand-crafted stack overlap in every repeat, which is consistent with a small effect that the paired tests detect in two of the three repeats. The paired tests are the more sensitive comparison here, because they use the fact that both models classify the same clips.

\begin{table}[!htbp]
\caption{Pooled test accuracy (\%) of each repeat with its 95\% bootstrap confidence interval (2,000 resamples; 2,664 clips per repeat, five folds pooled).}
\label{tab:bootstrap}
\tablesize{\normalsize}
\setlength{\tabcolsep}{5pt}
\renewcommand{\arraystretch}{1.3}
\begin{tabularx}{\textwidth}{Lccc}
\toprule
\textbf{Model} & \textbf{Repeat 1} & \textbf{Repeat 2} & \textbf{Repeat 3} \\
\midrule
RF, hybrid mRMR+\allowbreak{}ReliefF & 91.25 [90.17, 92.34] & 91.55 [90.43, 92.61] & 91.63 [90.61, 92.72] \\
\arrayrulecolor{black!25}\midrule[0.3pt]\arrayrulecolor{black}
XGBoost, hybrid-57 & 94.74 [93.88, 95.61] & 94.26 [93.36, 95.16] & 94.74 [93.88, 95.61] \\
\arrayrulecolor{black!25}\midrule[0.3pt]\arrayrulecolor{black}
Stacking: hand-crafted only & 94.41 [93.54, 95.27] & 94.11 [93.21, 95.01] & 94.18 [93.28, 95.08] \\
\arrayrulecolor{black!25}\midrule[0.3pt]\arrayrulecolor{black}
Stacking: hand-crafted + YAMNet (proposed) & 95.23 [94.41, 96.02] & 95.16 [94.33, 95.98] & 95.61 [94.78, 96.40] \\
\bottomrule
\end{tabularx}
\end{table}

Table~\ref{tab:ratio} compares the two-view stack trained on the balanced 1:1 pool with the class-weighted 1:3 pool. The balanced version reached a higher accuracy (95.97\%) and recall (95.48\%), while the class-weighted pool gave higher specificity (98.29\% against 96.46\%) and fewer false positives on unseen ESC-50 clips (98.78\% against 96.98\%). The training ratio thus sets the balance between missed drones and false alarms at the fixed 0.5 threshold, and its effect on specificity is larger than the effect of fusion itself.

\begin{table}[!htbp]
\caption{Effect of the training class ratio on the two-view stack: balanced 1:1 pool versus class-weighted 1:3 pool.}
\label{tab:ratio}
\tablesize{\small}
\setlength{\tabcolsep}{3pt}
\renewcommand{\arraystretch}{1.3}
\begin{tabularx}{\textwidth}{Lccccccc}
\toprule
\textbf{Model} & \textbf{Acc.} & \textbf{Prec.} & \textbf{Rec.} & \textbf{F1} & \textbf{Spec.} & \textbf{AUC} & \textbf{ESC-50 spec.} \\
\midrule
Stacking: hybrid of 44 features + YAMNet, balanced pool & 95.52 $\pm$ 1.06 & 95.83 $\pm$ 1.77 & 95.23 $\pm$ 2.10 & 95.50 $\pm$ 1.08 & 95.81 $\pm$ 1.83 & 0.988 $\pm$ 0.008 & 96.73 $\pm$ 0.34 \\
\arrayrulecolor{black!25}\midrule[0.3pt]\arrayrulecolor{black}
Stacking: hybrid of 57 features + YAMNet, balanced pool & 95.97 $\pm$ 1.70 & 96.45 $\pm$ 1.77 & 95.48 $\pm$ 2.79 & 95.94 $\pm$ 1.75 & 96.46 $\pm$ 1.81 & 0.990 $\pm$ 0.009 & 96.98 $\pm$ 0.35 \\
\arrayrulecolor{black!25}\midrule[0.3pt]\arrayrulecolor{black}
Stacking: hand-crafted only & 94.18 $\pm$ 2.10 & 97.68 $\pm$ 0.95 & 90.52 $\pm$ 4.35 & 93.91 $\pm$ 2.39 & 97.84 $\pm$ 0.94 & 0.988 $\pm$ 0.008 & 98.45 $\pm$ 0.18 \\
\arrayrulecolor{black!25}\midrule[0.3pt]\arrayrulecolor{black}
Stacking: hand-crafted + YAMNet (proposed) & 95.29 $\pm$ 1.62 & 98.19 $\pm$ 0.92 & 92.30 $\pm$ 3.14 & 95.12 $\pm$ 1.76 & 98.29 $\pm$ 0.89 & 0.990 $\pm$ 0.009 & 98.78 $\pm$ 0.10 \\
\bottomrule
\end{tabularx}
\par\smallskip{\footnotesize Mean $\pm$ SD over 15 grouped test folds, in percent (AUC as a fraction); ESC-50 spec.: unseen-source specificity, mean $\pm$ SD over 3 repeats.}
\end{table}

\subsection{Error analysis and computational cost}
\label{sec:res-esc}
\label{sec:res-latency}
\begin{figure}[!htbp]
\centering
\includegraphics[width=0.75\textwidth]{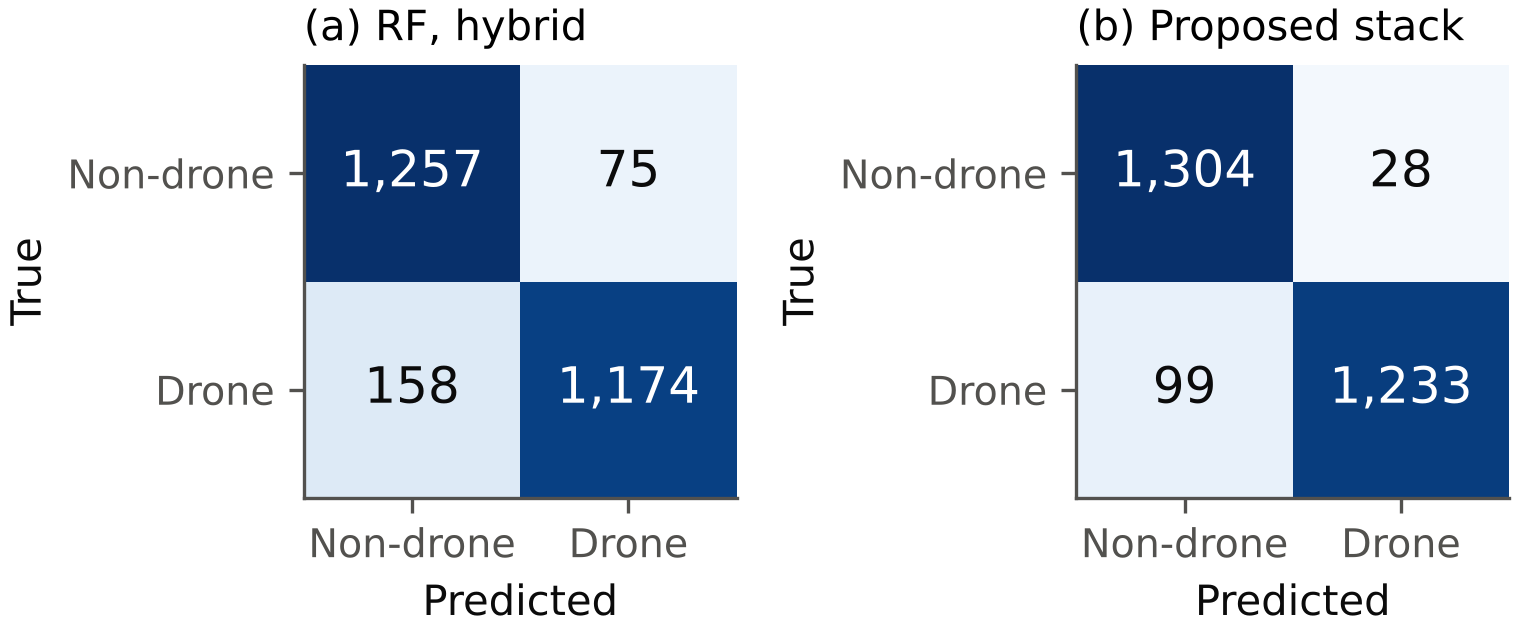}
\caption{Confusion matrices of the Random Forest baseline and the proposed stack, grouped protocol, first repeat (five folds pooled, 2,664 test clips).}
\label{fig:confusion}
\end{figure}

\begin{figure}[!htbp]
\centering
\includegraphics[width=0.7\textwidth]{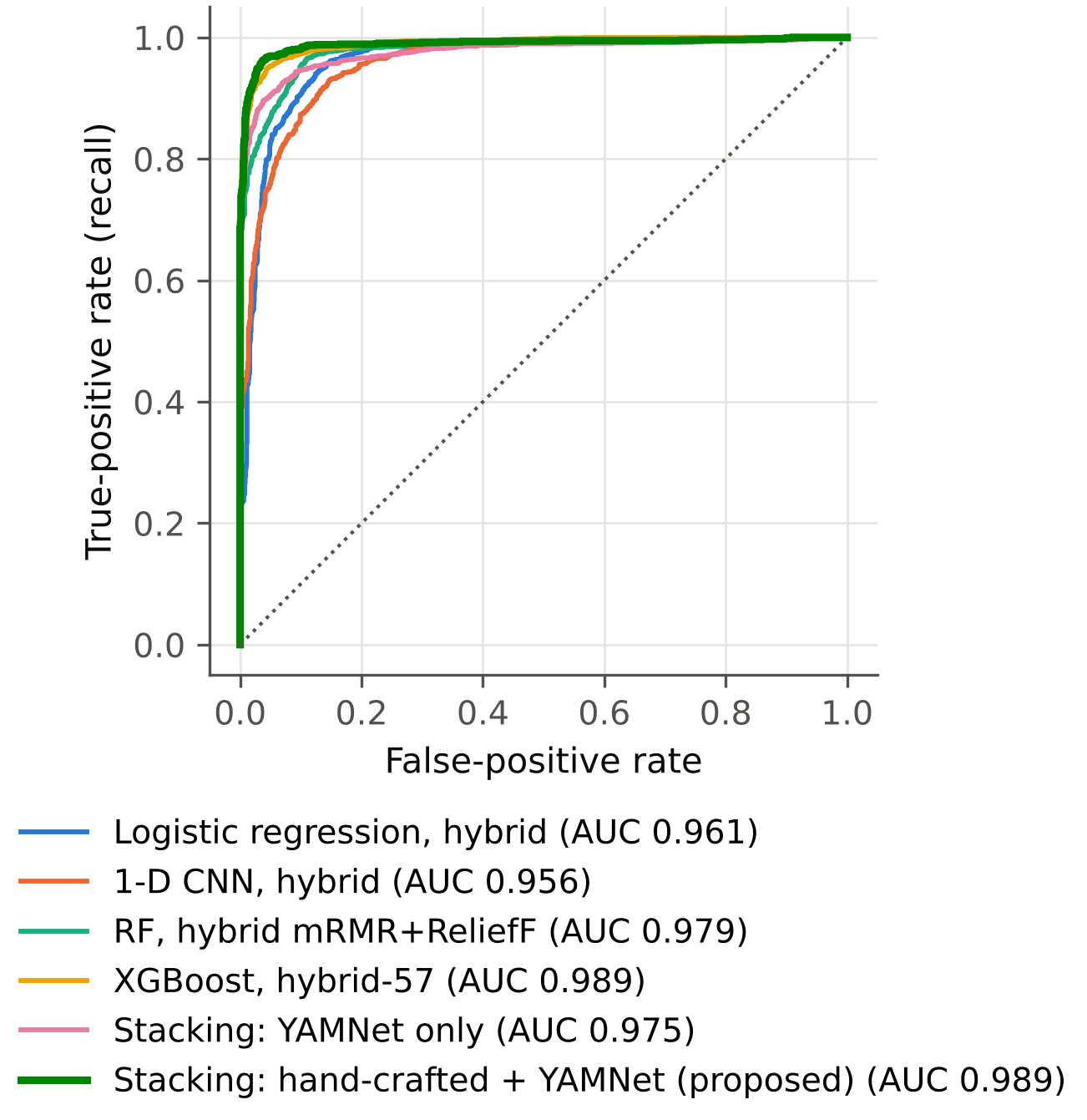}
\caption{ROC curves of representative models, grouped protocol, first repeat (five folds pooled).}
\label{fig:roc}
\end{figure}

Pooled over the five folds of the first repeat (Figure~\ref{fig:confusion}), the proposed stack produced 28 false positives and 99 missed drones among 2,664 test clips, fewer of both kinds than the Random Forest baseline. Missed drones outnumber false alarms, in line with the high specificity of the class-weighted pool. The ROC curve of the proposed stack lies at or above those of the other models at low false-positive rates (Figure~\ref{fig:roc}), the region that matters when false alarms are costly.

\begin{figure}[!htbp]
\centering
\includegraphics[width=0.75\textwidth]{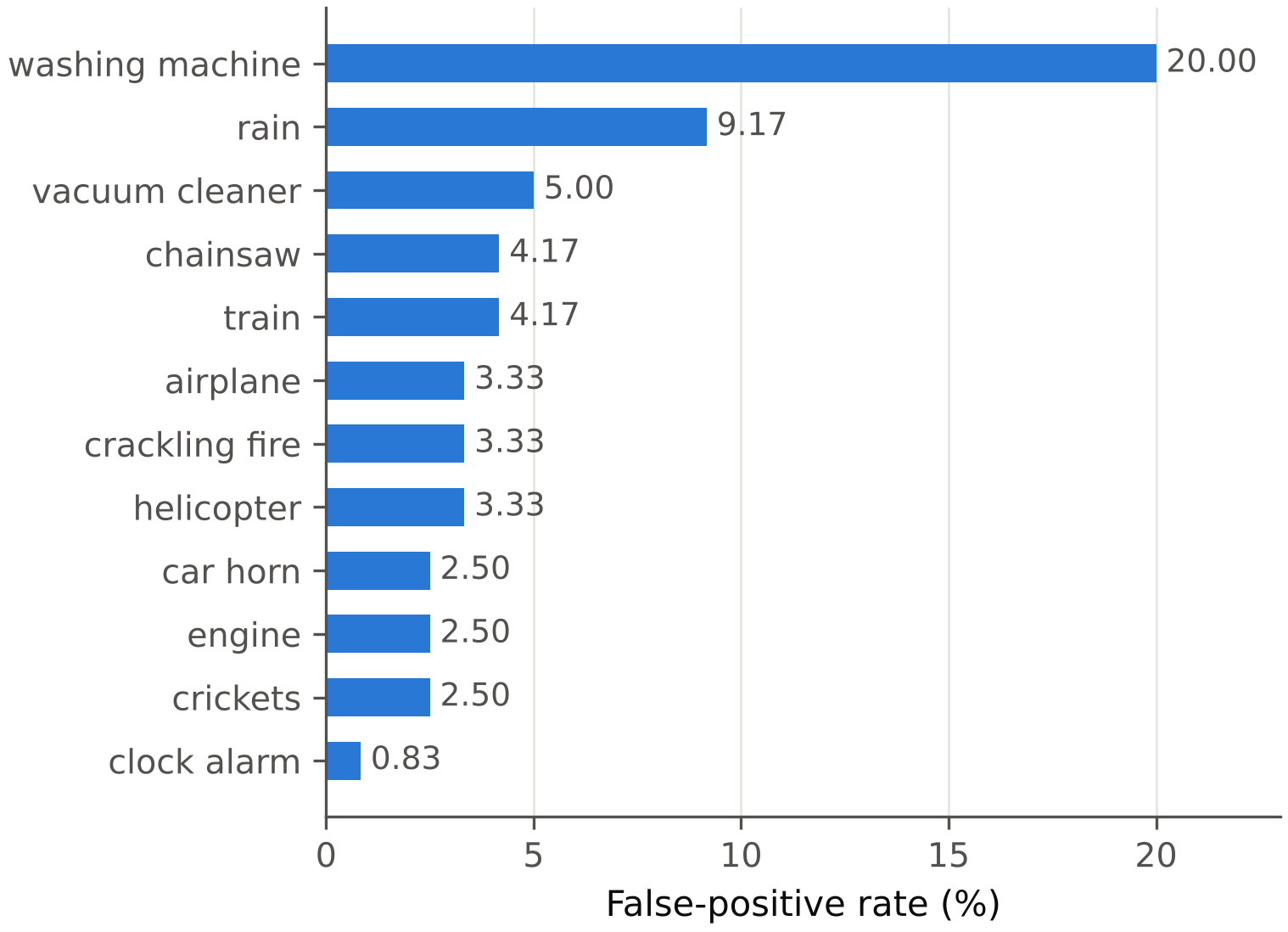}
\caption{ESC-50 classes most often scored as drone by the proposed stack (each clip scored by a model that never saw it; mean over 3 repeats).}
\label{fig:esc}
\end{figure}

On unseen ESC-50 clips (Table~\ref{tab:escclasses} and Figure~\ref{fig:esc}), the most frequent confusions were with steady broadband or motorised sounds: washing machine (20.00\%), rain (9.17\%), vacuum cleaner (5.00\%), and chainsaw and train (4.17\% each). Airplane and helicopter sounds were each mistaken for drones in 3.33\% of clips. By major sound group (Table~\ref{tab:escgroups}), no animal or human non-speech clip was labelled as a drone, and the interior/domestic group had the highest false-positive rate (2.58\%). The errors are concentrated in a few classes with sustained mechanical or noise-like sound rather than spread over the whole collection.

\begin{table}[!htbp]
\caption{ESC-50 classes most often scored as drone by the proposed stack.}
\label{tab:escclasses}
\tablesize{\normalsize}
\setlength{\tabcolsep}{5pt}
\renewcommand{\arraystretch}{1.3}
\begin{tabularx}{\textwidth}{LLc}
\toprule
\textbf{ESC-50 class} & \textbf{Major group} & \textbf{False-positive rate (\%), mean $\pm$ SD} \\
\midrule
washing machine & Interior/domestic & 20.00 $\pm$ 2.50 \\
\arrayrulecolor{black!25}\midrule[0.3pt]\arrayrulecolor{black}
rain & Natural soundscapes & 9.17 $\pm$ 3.82 \\
\arrayrulecolor{black!25}\midrule[0.3pt]\arrayrulecolor{black}
vacuum cleaner & Interior/domestic & 5.00 $\pm$ 0.00 \\
\arrayrulecolor{black!25}\midrule[0.3pt]\arrayrulecolor{black}
chainsaw & Exterior/urban & 4.17 $\pm$ 1.44 \\
\arrayrulecolor{black!25}\midrule[0.3pt]\arrayrulecolor{black}
train & Exterior/urban & 4.17 $\pm$ 1.44 \\
\arrayrulecolor{black!25}\midrule[0.3pt]\arrayrulecolor{black}
airplane & Exterior/urban & 3.33 $\pm$ 1.44 \\
\arrayrulecolor{black!25}\midrule[0.3pt]\arrayrulecolor{black}
crackling fire & Natural soundscapes & 3.33 $\pm$ 1.44 \\
\arrayrulecolor{black!25}\midrule[0.3pt]\arrayrulecolor{black}
helicopter & Exterior/urban & 3.33 $\pm$ 1.44 \\
\arrayrulecolor{black!25}\midrule[0.3pt]\arrayrulecolor{black}
car horn & Exterior/urban & 2.50 $\pm$ 0.00 \\
\arrayrulecolor{black!25}\midrule[0.3pt]\arrayrulecolor{black}
engine & Exterior/urban & 2.50 $\pm$ 2.50 \\
\bottomrule
\end{tabularx}
\par\smallskip{\footnotesize Mean $\pm$ SD over 3 repeats; 40 source clips per class.}
\end{table}
\begin{table}[!htbp]
\caption{ESC-50 false-positive rate of the proposed stack by major sound group (400 source clips per group, mean over 3 repeats).}
\label{tab:escgroups}
\tablesize{\normalsize}
\setlength{\tabcolsep}{5pt}
\renewcommand{\arraystretch}{1.3}
\begin{tabularx}{\textwidth}{Lc}
\toprule
\textbf{Major group} & \textbf{False-positive rate (\%)} \\
\midrule
Animals & 0.00 \\
\arrayrulecolor{black!25}\midrule[0.3pt]\arrayrulecolor{black}
Exterior/urban & 2.00 \\
\arrayrulecolor{black!25}\midrule[0.3pt]\arrayrulecolor{black}
Human non-speech & 0.00 \\
\arrayrulecolor{black!25}\midrule[0.3pt]\arrayrulecolor{black}
Interior/domestic & 2.58 \\
\arrayrulecolor{black!25}\midrule[0.3pt]\arrayrulecolor{black}
Natural soundscapes & 1.50 \\
\bottomrule
\end{tabularx}
\end{table}

\begin{figure}[!htbp]
\centering
\includegraphics[width=0.8\textwidth]{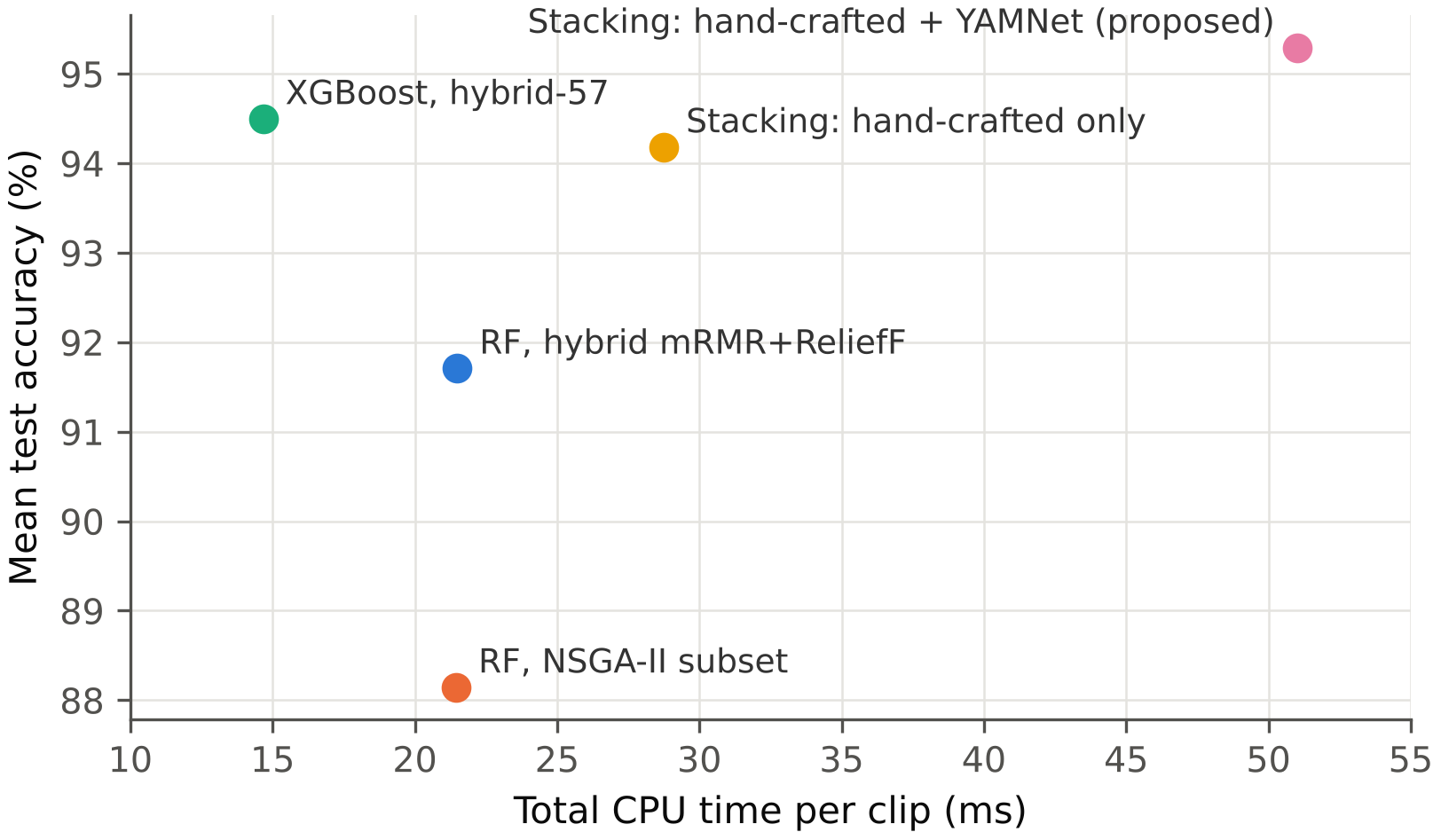}
\caption{Accuracy against total single-thread CPU time per clip. The NSGA-II point averages the five folds of the first repeat; the others average 15 folds.}
\label{fig:latency}
\end{figure}

Table~\ref{tab:latency} and Figure~\ref{fig:latency} report single-thread CPU cost. Extracting the 57 descriptors took 14.4~ms per clip and YAMNet 7.7~ms. XGBoost on the hand-crafted subset needed 14.7~ms per clip in total with a 0.69~MB model, the hand-crafted stack 28.8~ms, and the two-view stack 51.0~ms with 12.6~MB of classifiers. For the single XGBoost model, feature extraction accounts for almost all of the time; for the two-view stack, the four base learners and the meta-learner take more than half. All pipelines process a one-second clip well within one second on a single CPU thread, which leaves a wide margin for continuous operation on the processor used here.

\begin{table}[!htbp]
\caption{Single-thread CPU cost per 1-s clip on an AMD EPYC 7513 32-Core Processor (median of 100 clips $\times$ 3 repeats).}
\label{tab:latency}
\tablesize{\small}
\setlength{\tabcolsep}{3pt}
\renewcommand{\arraystretch}{1.3}
\begin{tabularx}{\textwidth}{Lccccccc}
\toprule
\textbf{Pipeline} & \textbf{Hand-crafted features} & \textbf{YAMNet} & \textbf{Features (ms)} & \textbf{Classifier (ms)} & \textbf{Total (ms)} & \textbf{Size (MB)} & \textbf{Accuracy (\%)} \\
\midrule
RF, hybrid mRMR+\allowbreak{}ReliefF & 17 & no & 14.43 & 7.06 & 21.49 & 1.68 & \begin{tabular}[c]{@{}c@{}}91.71\\$\pm$ 3.30\end{tabular} \\
\arrayrulecolor{black!25}\midrule[0.3pt]\arrayrulecolor{black}
RF, NSGA-II subset & 6 & no & 14.43 & 7.02 & 21.45 & 2.15 & \begin{tabular}[c]{@{}c@{}}88.14\\$\pm$ 2.03$^{a}$\end{tabular} \\
\arrayrulecolor{black!25}\midrule[0.3pt]\arrayrulecolor{black}
XGBoost, hybrid of 57 features & 18 & no & 14.43 & 0.24 & 14.67 & 0.69 & \begin{tabular}[c]{@{}c@{}}94.50\\$\pm$ 2.24\end{tabular} \\
\arrayrulecolor{black!25}\midrule[0.3pt]\arrayrulecolor{black}
Stacking: hand-crafted only & 18 & no & 14.43 & 14.32 & 28.75 & 5.44 & \begin{tabular}[c]{@{}c@{}}94.18\\$\pm$ 2.10\end{tabular} \\
\arrayrulecolor{black!25}\midrule[0.3pt]\arrayrulecolor{black}
Stacking: hand-crafted + YAMNet (proposed) & 18 & yes & 22.14 & 28.86 & 51.00 & 12.60 & \begin{tabular}[c]{@{}c@{}}95.29\\$\pm$ 1.62\end{tabular} \\
\bottomrule
\end{tabularx}
\par\smallskip{\footnotesize Accuracy: grouped protocol, mean $\pm$ SD over 15 folds; $^{a}$ five folds of the first repeat.}
\end{table}

\section{Discussion}

The protocol comparison shows that a random split of the DroneAudioDataset rewards memorising recordings and source sounds. The inflation of 1.89 to 3.22 percentage points is of the same order as the differences between the models compared here, so accuracies obtained with random splits on this dataset cannot be compared with source-disjoint results or taken as evidence of generalisation. The grouped protocol also exposes much larger fold-to-fold variation, which a single split hides. These findings are consistent with Shulman's grouped evaluation of the same dataset~\cite{shulman2026}. Because most non-drone clips derive from ESC-50, ESC-50 is also unsuitable as an external test set for models trained on this dataset unless its source clips are excluded from training. For future benchmarks, the recording and source identifiers already present in the file names make a grouped split easy to apply, and reporting the grouping, the number of folds and the fold assignment would let results from different studies be compared directly.

Hybrid mRMR+ReliefF selection kept the accuracy of the full descriptor set with fewer than half of the features, and the descriptors chosen in every fold (low-order MFCC means, spectral bandwidth, spectral contrast and the spread of the log-mel spectrogram) are consistent with the broadband, harmonic character of rotor noise. Because the selected descriptors came from every family, a detector built on MFCCs alone would likely discard useful information. NSGA-II offers a different operating point: four to six descriptors at a few percentage points lower accuracy, which suits settings where feature computation or memory is tightly limited.

Fusion with YAMNet embeddings gave a small, consistent improvement over the best hand-crafted models. The YAMNet view alone was weaker than the hand-crafted view, yet the meta-learner relied on it almost as much as on the hand-crafted Random Forest, which suggests that the two views make partly different errors. The improvement appeared mainly as higher recall, so the embedding mainly recovers drone clips that the hand-crafted view misses. The gain over XGBoost on the hand-crafted subset was below one standard deviation of the fold accuracies and was significant in two of the three repeats, so it should be read as a modest benefit. The class-ratio comparison shows that the training pool, more than the fusion itself, controls the balance between recall and specificity; applications that must avoid false alarms favour the class-weighted pool, while applications that must not miss a drone favour the balanced pool or a lower threshold.

The ESC-50 error analysis indicates where acoustic detectors trained on this dataset are fragile: steady machine and broadband noises such as washing machines, vacuum cleaners and rain cause most false alarms, while aircraft sounds are rarely confused. This information is useful for choosing negative training material and for planning field tests. A detector intended for a particular site would benefit from negative recordings of the machines and weather typical of that site, and a field evaluation should include them on purpose rather than rely on random background sound.

For a fixed acoustic sensor, the results describe a set of operating points rather than a single best model. The single XGBoost model on the hand-crafted subset is the cheapest option with competitive accuracy, the two-view stack adds 0.80 percentage points of accuracy and higher recall at more than three times the CPU time, and an NSGA-II subset reduces the descriptor set further for the most constrained devices. All of these choices remain fast enough for real-time processing of one-second clips on a single core of the processor used in this study.

\section{Conclusions}

Under a grouped, source-disjoint protocol for the DroneAudioDataset binary task, a stacking ensemble of hand-crafted descriptors and pretrained YAMNet embeddings reached 95.29\% accuracy and 98.29\% specificity, and processed a clip in 51.0~ms on one CPU thread. Random clip-level splitting overstated accuracy by 1.89 to 3.22 percentage points, a margin comparable to the differences between models, so source-disjoint evaluation should be the default for this benchmark. Hybrid mRMR+ReliefF selection matched the full descriptor set with fewer features, NSGA-II provided compact subsets for constrained settings, and fusion with pretrained embeddings added a small but consistent gain, mainly through higher recall. The unseen-source analysis showed that the remaining false alarms come from a few steady mechanical and broadband sounds, which identifies the negative material that future training sets should contain. Future work will test the models on independently recorded outdoor data with more drone types, evaluate further pretrained and self-supervised audio encoders, and study federated training of the detector across distributed sensor nodes~\cite{iqbal2021fl}, where differences in data and model capacity between nodes can be handled by confidence-weighted knowledge distillation~\cite{ma2021,iqbal2026}.

\end{document}